\documentstyle[pra,aps]{revtex}
\twocolumn
\begin{document}
\draft
\title{CLASSICAL AND QUANTUM MALUS' LAW}
\author{Krzysztof W\'odkiewicz}
\address{Center for Advanced Studies Department of Physics and
Astronomy, \\
University of New Mexico, Albuquerque, New Mexico 87131, USA \\
and \\
Instytut Fizyki Teoretycznej, Uniwersytet Warszawski,\\
Warszawa 00-681, Ho\.za 69, Poland \cite{aaa}}
\date{\today}
\maketitle
\begin{abstract}
The classical and the quantum Malus' Laws for  light and spin are discussed.
It is shown that for spin-$\frac{1}{2}$, the quantum Malus' Law is
equivalent in form to the classical Malus' Law provided that  the
statistical average involves a quasi-distribution function that can become
negative. A  generalization of  Malus' Law for  arbitrary spin-$s$ is
obtained in the form of a Feynman path-integral representation for the Malus
amplitude. The classical limit of the Malus amplitude for ${s\to \infty}$
is discussed.

\end{abstract}



\section{Introduction}

The classical Malus' Law predicts an attenuation of a polarized
light beam through a linear polarizer.  This attenuation depends on the
relative angle $\alpha$ between the polarization direction $\vec a$ of
the incoming wave and  the orientation ${\vec a}^{\prime}$ of the polarizer.
According to Malus' Law the attenuation of the light intensity is just
$\cos^{2} \alpha$. If the incoming beam consists of a statistical mixture
of polarized
light,  the probability to go through a linear polarizer is:

\begin{equation}\label{cmalus}
p = \int d \Omega P_{\it cl}(\Omega) \cos^{2} \alpha .
\end{equation}%
In this formula the integration is over all possible angles  of the random
polarization
direction $\vec a$  described  by a solid angle $\Omega = (\theta, \phi)$
and where the classical distribution function $P_{\it cl}(\Omega)$
characterizes the statistical properties of the   incident light beam
polarization.

It is the purpose of this paper to discuss the quantum Malus' Law for
spin systems. Entangled spin correlations provide examples of such systems. We
show that in general, the quantum Malus' Law is equivalent in form to the
classical Malus' Law provided that  the statistical average involves a
quasi-distribution function that can become negative. A generalization of the
Malus' Law for an arbitrary spin-$s$ system is obtained. Using a Feynman
path-integral representation for
the Malus' amplitude the relation between Malus' amplitude and the Malus' Law
is obtained in the limit of  ${s\to \infty}$. The classical limit of the Malus'
Law is discussed, and classical equations of motion for spin systems are
derived. The motivation for the discussion of the Malus' Law for higher spins
comes from the fact that systems  involving  higher-spin states
or many particles  exhibits very strong quantum correlations
\cite{mermin,zeilinger}.

\section{Quantum Malus' Law}

In quantum mechanics a similar Malus' Law holds for spin-$\frac{1}{2}$
particles detected by a Stern-Gerlach apparatus oriented in
the  direction $|\vec a^{\prime}\rangle$.
An arbitrary pure state of the spin-$\frac{1}{2}$ can be written as a
linear superposition of the up $|+\rangle$ and down $|-\rangle$ spin states:

\begin{equation}\label{omega}
|\Omega \rangle =e^{i \phi} \sin \frac{\theta}{2}|+\rangle + \cos
\frac{\theta}{2}|-
\rangle
\end{equation}
where the  solid angle $\Omega $ characterizes the
 spin orientation on a unit sphere (the Bloch sphere).

The quantum amplitude for the
transmission of a such a spin state  through a Stern-Gerlach apparatus is:
${\cal A}= \langle \Omega|\vec a^{\prime} \rangle$,
and  the probability is just the quantum Malus' transmission function:

\begin{equation}\label{malusp}
p = |{\cal A}|^{2} = \cos^{2} \frac{\alpha}{2}
\end{equation}
where

\begin{equation}
\cos \alpha = \cos \theta \cos \theta^{\prime} + \sin \theta
\sin \theta^{\prime} \cos (\phi -
\phi^{\prime})
\end{equation}
is the relative  angle between the spherical orientation $\Omega $
of the detected state (\ref{omega})  and the spherical direction
$\Omega^{\prime} $ of the Stern-Gerlach polarizer.  Photons and spins differ
in this formulation by a factor $\frac{1}{2}$ in the relative angle
involved in the Malus law.

Following the classical Malus' Law for an unpolarized light beam
(\ref{cmalus}),
one can write the following probability for an arbitrary  mixed state of the
spin-$\frac{1}{2}$ system detected by the Stern-Gerlach apparatus:

\begin{equation}\label{malus}
p = \int d \Omega P(\Omega) \cos^{2} \frac{\alpha}{2}.
\end{equation}
The  function $P(\Omega)$ plays the role of a statistical
distribution for an arbitrary beam of spin-$\frac{1}{2}$ particles.
In quantum mechanics one deals with probability amplitudes rather than
probabilities and one should sum these amplitudes first, before squaring
the result. Malus' amplitudes can be derived in such a way \cite{kw}.
Nevertheless the  formula (\ref{malus}) is valid in quantum mechanics if
the quantum mechanical distribution $P(\Omega)$ is  a quasi-distribution.
The quantum quasi-distribution function is associated with an arbitrary
density matrix $\hat {\rho}$ of the
spin-$\frac{1}{2}$ system in the following  way:
\begin{equation}
\hat {\rho} = \int d\Omega P(\Omega) |\Omega \rangle\langle \Omega |.
\end{equation}
In this expression  the diagonal weight function $P(\Omega)$ is a quantum
 quasi-probability distribution, and accordingly contains all the
statistical information about the spin state. This formula
is similar in its structure to the diagonal Glauber $P$--representation
for a harmonic oscillator if coherent states are used \cite{klauder}. For
spin-$\frac{1}{2}$ the corresponding spin coherent states (SCS) are  given by
unit directions (\ref{omega}) on the Bloch sphere. From the properties of
the SCS, one concludes that the quasi-distribution function is normalized
$\int d \Omega P(\Omega) = 1$, but in general it is not
positive definite or unique \cite{lieb}. For example, for the up and the down
spin states $|\pm \rangle$, the corresponding quasi-distributions are
$ P_{\pm}(\Omega)= \frac{1}{4\pi}(1\mp 3 \cos \theta)$. These functions
can take up negative values that indicate the  quantum character of
these states. As another example,  the
incoherent (mixed) state of the spin system described by the density matrix
$\hat{\rho} = \frac{1}{2} | +\rangle \langle +|+ \frac{1}{2} |
-\rangle \langle -|$,
leads to a  distribution that has a purely classical behavior corresponding
to a uniform distribution of directions on the Bloch sphere, i.e.,
$ P(\Omega)= \frac{1}{4\pi}$\cite{scully+kw1}.

The quantum character of the negative quasi-probability is seen best
for correlations involving a density operator for an entangled Einstein,
Podolsky, and Rosen state \cite{epr}. For such a correlated system of two
spin-$\frac{1}{2}$ particles, labelled by indices $a$ and $b$, the  singlet
wave function is:
\begin{equation}
|\psi\rangle = { 1 \over \sqrt{2}} \left(| + \rangle_a \otimes |-\rangle_b
- | - \rangle_a \otimes |+\rangle_b \right).
\end{equation}
The corresponding quasi-distribution function has the following form
\cite{scully+kw2}
\begin{equation}\label{pro1}
P(\Omega_{a};\Omega_{b}) =\frac{1}{(4\pi)^{2}}(1+9\cos \theta_{a} \cos
\theta_{b} +9 \sin \theta_{a}
\sin \theta_{b} \cos (\phi_{a} -
\phi_{b}).
\end{equation}

The quantum Malus Law (\ref{malus}), if applied to the
joint  correlations involving two Stern-Gerlach detectors (with directions
$\vec a$ and $\vec b$), provides the following joint probability for detection:
\begin{equation}\label{qmalus}
 p(\vec a; \vec b) = \int d
\Omega_a^\prime \int d \Omega^\prime_b P(\Omega^\prime_a; \Omega_b^\prime)
\cos^2\alpha(\Omega_a ,\Omega^\prime_a)
\cos^2 \alpha(\Omega_b,\Omega^\prime_b ) .
\end{equation}
This correlation function evaluated with the expression (\ref{pro1}) leads to
the well know quantum mechanical result for the joint spin correlation:
$ p(\vec a; \vec b)= \frac{1} {2} (1- {\vec a}\cdot {\vec b})$. The
distribution function (\ref{pro1}) is not unique, because the same result is
reproduced  if one  uses the following quasi-distribution function:
 \begin{equation}\label{pro2}
P(\Omega^\prime_a; \Omega^\prime_b) =
\frac{3}{4\pi}\delta^{(2)}(\Omega^\prime_a +\Omega^\prime_b) -
 \frac {2} { (4\pi)^2}.
\end{equation}
The formula (\ref{qmalus})  has the formal
structure of a hidden variable theory. In such a theory the joint probability
function is calculated from the expression:
\begin{equation}
p(\vec a;\vec b) =\int d\lambda_a \int d\lambda_b \ P(\lambda_a;\lambda_b)
\  t(\vec a,\lambda_a) \ t(\vec b,\lambda_b)
\end{equation}
where  $P(\lambda_a;\lambda_b)$ describes the
distribution of some  hidden variables
$\lambda_a$ and $\lambda_b$, and  the objective realities of the spin variables
are given by the deterministic transmission functions $t(\vec a,\lambda_a)$ and
$ t(\vec b,\lambda_b)$ through the Stern-Gerlach apparatus.  It is  clear that
the quantum mechanical formula (\ref{qmalus})
has the form of a hidden variable theory with the local spin realities given by
$\cos^2\alpha(\Omega_a ,\Omega^\prime_a)$ and $\cos^2
\alpha(\Omega_b,\Omega^\prime_b ) $. In such a theory  the hidden parameters
are represented by "hidden angles" on a Bloch sphere, and are distributed
according to Eq.~(\ref{pro1}) or Eq.~(\ref{pro2}).  There the analogy ends
because the
quantum distribution of these "hidden directions" is given by a non-positive
function that leads to the failure of the Bell's inequalities for
such an entangled state.

The SCS and the quantum Malus' Law (\ref{malus}) can be generalized to an
arbitrary spin-$s$. The spin-$s$ coherent states are
obtained by a rotation of the maximum "down" spin state $|s,-s\rangle$
\cite{arecchi}:
\begin{equation}
|\Omega \rangle =\exp( \tau {\hat S}_{+} -
\tau^{*} {\hat S}_{-})|s,-s\rangle ,
\end{equation}
where $\tau=\frac{1}{2}\theta e^{-i\phi}$ and   ${\hat S}_{\pm}$ are
the spin-$s$ ladder operators.
The SCS form an over complete set of states on the Bloch
sphere:

\begin{equation}
\frac{2s+1}{4\pi} \int d \Omega
 |\Omega\rangle \langle \Omega|= I.
\end{equation}
Using these formulas, it is easy to calculate the
Malus' quantum amplitude and the probability for a transmission of
such a  state through a Stern-Gerlach apparatus. As a result one obtains:

\begin{equation}
p =|\langle \Omega|\Omega^{\prime}\rangle|^{2} =
(\cos \frac{\alpha}{2})^{4s} ,
\end{equation}
with a straightforward generalization involving an arbitrary
quasi-distribution
function $P(\Omega)$ for a density operator of a system with arbitrary
spin-$s$. This quantum mechanical expression for
the transmission function  provides a generalization of the
spin-$\frac{1}{2}$ Malus' Law (\ref{malusp}) to the case of an arbitrary
spin-$s$ system.

\section{Path-integral form of Malus' Law}

This quantum Malus Law for arbitrary spin, is well suited to study the
relation between classical and quantum features of the transmission
function.
In quantum mechanics the primary object is the probability amplitude for the
transmission of a SCS $|\Omega \rangle$ through a
Stern-Gerlach apparatus $|\vec a \rangle$
characterized by a solid angle $\Omega^{\prime} $. This probability amplitude
is
${\cal A}= \langle \Omega|\Omega^{\prime}\rangle$,
and can be cast into a path-integral form exhibiting various quantum paths
contributing to the transition. Following the basic idea of
 path integration \cite{feynman}, one can evaluate the Malus amplitude by
dividing the spin trajectories on the Bloch sphere into infinitesimal
subintervals
$|\Omega_{i} \rangle$
where $i=1, \ldots , N$ with $\Omega_{1} = \Omega^{\prime}$ and
$\Omega_{N} = \Omega$.
Using the decomposition of unity  for the SCS for each
subinterval and the infinitesimal form of the Malus amplitude
$\langle \Omega_{i}|\Omega_{i-1} \rangle $, one obtains

\begin{eqnarray}
{\cal A} = \int d \Omega_{1} \frac{2s+1}{4\pi}
\int d \Omega_{2} \frac{2s+1}{4\pi}, \ldots \nonumber\\
\exp\left(-is \sum_i (\phi_i - \phi_{i-1})\cos\theta_{i-1} \right).
\end{eqnarray}
In the limit of $N\to \infty$,  this expression can be written in
the form of the spin Feynman path-integral:

\begin{equation}\label{path}
{\cal A} = \int {\cal D} \Omega \frac{2s+1}{4\pi}
\exp(-is \int d \phi \cos\theta )
\end{equation}
where ${\cal D} \Omega$ is the functional path-integration
measure over all spin trajectories connecting $|\Omega^{\prime} \rangle$
with
$|\Omega \rangle$ on the Bloch sphere. This path-integral representation
of the quantum Malus Law can be cast in a more familiar form if
the spherical angles are identified with the canonical position and
the canonical momentum in the following way:  $\phi \Leftrightarrow q, \;
\cos \theta \Leftrightarrow p$, and $d\Omega=d \phi d \cos \theta
\Leftrightarrow   dq dp $. Using
these variables we can rewrite the path-integral (\ref{path}) in the form:

\begin{equation}
{\cal A} =
\int {\cal D}q  {\cal D}p   \frac{2s+1}{4\pi} \exp(-is \int pdq )
\end{equation}
which is the spin analog of the phase-space path integral for the
following quantum mechanical amplitude in the configuration space:
\begin{equation}
\langle q|q^{\prime} \rangle = \int \frac{{\cal D}q {\cal D}p}
{ 2 \pi \hbar}
\exp(-{i\over \hbar} \int dq p).
\end{equation}

The Malus probability for the spin-$s$ transition of the state
$|\Omega\rangle$ through  such a Stern-Gerlach apparatus can be
expressed as a product of four path integrals:

\begin{eqnarray}
|{\cal A}|^{2} = \int {\cal D}q_{1}  {\cal D}p_{1}   \frac{2s+1}{4\pi}
\int {\cal D}q_{2} {\cal D}p_{2}  \frac{2s+1}{4\pi} \nonumber\\
\exp\left( i{\cal S}(q_{1},p_{1}) -i{\cal S}(q_{2},p_{2})\right),
\end{eqnarray}
where the classical  action is

\begin{equation}
{\cal S}(q,p)=s \int p dq + {1\over \hbar} \int {\cal H} dt .
\end{equation}
In this expression a classical Hamiltonian ${\cal H}$ has been added in order
to describe a  possible dynamical time evolution before the particle
has reached the Stern-Gerlach apparatus.

In order to see the connection with the quantum Malus Law (\ref{malus}) and
the classical
Malus Law (\ref{cmalus}) one can investigate the properties of the Malus'
transmission
function in the classical limit corresponding to $s \to \infty$.
The transition from quantum amplitudes to classical probabilities can be
carried out, if the four path-integrals can be simplified. In configuration
space the classical limit  of the path-integral can be investigated using
a suitable change of variables \cite{birula}. In the case of the
path-integral for the Malus' probability this change of variables is:

\begin{eqnarray}
q_{1,2} = q \pm {1\over 2s} \tilde q, \nonumber \\
p_{1,2} = p \pm {1\over 2s} \tilde p.
\end{eqnarray}
In the limit of $s \to \infty$,  the path-integrals with respect
to ${\cal D}\tilde q$ and $ {\cal D}\tilde p$
can be performed, leading to functional Dirac's functions, and the entire
expression for the probability simplifies to:

\begin{equation}\label{classical}
\lim_{s\to \infty} |{\cal A}|^2 = \int {\cal D}q \int {\cal D}p \
\delta (\dot q -{1\over s\hbar} {\partial {\cal H}\over \partial p}) \
\delta (\dot p +{1\over s\hbar} {\partial {\cal H}\over \partial q}).
\end{equation}
This expression shows that in the classical limit, the spin-$s$ Malus
transmission function reduces to a classical dynamics on the Bloch sphere
with the  following canonical equations of motion:

\begin{equation}
\dot \phi = \lbrace \phi,{\cal H}\rbrace \ \ {\rm and} \ \
 \dot \theta = \lbrace \theta,{\cal H}\rbrace.
\end{equation}
{}From the reduced path-integral formula (\ref{classical}) one obtains that the
Poisson bracket
of the classical dynamics is:
\begin{equation}
\lbrace A,B \rbrace = {1\over s\hbar \sin \theta} (
{\partial A \over \partial \phi}  {\partial B \over \partial \theta}-
{\partial A \over \partial \theta} {\partial B \over \partial \phi}).
\end{equation}
In these equations one recognizes the classical equations of motion of a
particle confined to a sphere. The Poisson bracket in this case has a typical
structure for a curved phase-space associated with the Bloch sphere
\cite{berezin}.

If the Malus' law is applied to an arbitrary spin-$s$ system described by
a  quasi-distribution function, in the limit $s \to \infty$
the expression (\ref{classical}) corresponds to a classical statistical
mechanics  on a unite sphere. These classical trajectories are distributed
with a classical distribution function $P_{cl}(\Omega)$  emerging from
$P(\Omega)$  in the limit $s \to \infty$.

\section*{Acknowledgments}

The author thanks C. Caves and G. Herling for
numerous discussions and comments.
This work was partially supported by the Polish KBN
Grant No. 20 426 91 01, and the Center for Advanced Studies of the
University of New Mexico.


\begin{references}

\bibitem[*]{aaa}
Permanent address


\bibitem{mermin}
N. D. Mermin, Phys. Rev. D {\bf 22}, 356 (1980).

\bibitem{zeilinger}
 M. A. Horne, A. Shimony, and A. Zeilinger, Am. J. Phys. {\bf 58}, 1131 (1990).

\bibitem{kw}
K. W\'odkiewicz, Phys. Lett. {\bf 112A} (1986) 304.

\bibitem{klauder}
See, for example, J. R. Klauder and B. S. Skagerstam, {\it Coherent
States} (World Scientific, Singapore, 1985).

\bibitem{lieb}
E. H. Lieb, Commun. Math. Phys. {\bf31}, 327 (1973).

\bibitem{scully+kw1} For various spin quasi-distributions see, for example,
M.~O.~Scully and K.~W\'odkiewicz
 Found. of Phys. {\bf 24} (1994) 85.




\bibitem{epr} A. Einstein, B. Podolsky, and N. Rosen, Phys. Rev. {\bf 47}, 777
(1935).




\bibitem{scully+kw2}
 M. O. Scully and K. W\'odkiewicz, {\bf Coherence and  Quantum Optics VI},
eds. J.~H.~Eberly et al. (Plenum, New York, 1990) p. 1047.


\bibitem{arecchi}
F. T. Arrechi, E. Courtens, R. Gilmore and H. Thomas, Phys. Rev. A {\bf
6} (1972) 2211.
\bibitem{feynman}
R. P. Feynman and R. Hibbs, {\bf Quantum Mechanics and path integrals}
(McGrowHill, New York, 1965).


\bibitem{birula}
I. Bia\l ynicki-Birula Ann. Phys. {\bf 67} (1971) 252.
\bibitem{berezin}
F. A. Berezin, {\bf The Method of Second Quantization} (Academic, New York,
1966).
\end{references}
\end{document}